\documentclass{elsart}

\usepackage{graphicx}
\usepackage{epsfig}

\usepackage{amssymb}

\begin{document}
\def \gr{$\gamma$-ray }
\def\st{\scriptstyle}
\def\sst{\scriptscriptstyle}
\def\mco{\multicolumn}
\def\epp{\epsilon^{\prime}}
\def\vep{\varepsilon}
\def\ra{\rightarrow}
\def\ppg{\pi^+\pi^-\gamma}
\def\vp{{\bf p}}
\def\ko{K^0}
\def\kb{\bar{K^0}}
\def\al{\alpha}
\def\cal{\mathcal}
\def\ab{\bar{\alpha}}
\def\be{\begin{equation}}
\def\ee{\end{equation}}
\def\bea{\begin{eqnarray}}
\def\eea{\end{eqnarray}}
\def\CPbar{\hbox{{\rm CP}\hskip-1.80em{/}}}
\def\gsim{\;\lower4pt\hbox{${\buildrel\displaystyle >\over\sim}$}\,}
\def\lsim{\;\lower4pt\hbox{${\buildrel\displaystyle <\over\sim}$}\,}
\def \sax {BeppoSAX}
\def \xmm {{\em XMM-Newton}}
\def \degmark{^\circ}
\def \nh {N${\rm _H}$}
\def \ergsec{\hbox{erg s$^{-1}$}}
\def \hcm {\hbox {\ifmmode $ atom cm$^{-2}\else atom cm$^{-2}$\fi}}
\def \arcmin {\hbox{$^\prime$}}
\def \arcsec {\hbox{$^{\prime\prime}$}}
\def \chisq {$\chi ^{2}$}

\def \rchisq {$\chi_{\nu} ^{2}$}
\def\approxgt{\mathrel{\hbox{\rlap{\lower.55ex \hbox {$\sim$}}
        \kern-.3em \raise.4ex \hbox{$>$}}}}
\def\approxlt{\mathrel{\hbox{\rlap{\lower.55ex \hbox {$\sim$}}
        \kern-.3em \raise.4ex \hbox{$<$}}}}
\newcommand{\mc}{\multicolumn}
\newcommand {\Msun}{M_\odot}
\def\CPbar{\hbox{{\rm CP}\hskip-1.80em{/}}}
\def\gcm{\rm ~g~cm^{-3}}
\def\cmc{\rm ~cm^{-3}}
\def\cms{\rm ~cm^{-2}}
\def \cmsec {\rm ~cm~s^{-1}}
\def\kms{\rm ~km~s^{-1}}
\def\ergs{\rm ~erg~s^{-1}}
\def\enf{\rm ~erg~s^{-1}~cm^{-2}}
\def\ml{~\Msun ~\rm yr^{-1}}
\def\mll{\Msun ~\rm yr^{-1}}
\def\arcs{$^{\prime\prime}$}
\def\arcm{$^{\prime}$}
\def\etal{{et al.}}
\def\alf{Alfv\'en~}
\def\bs{{\it BeppoSAX~~}}
\def \sax {{\it BeppoSAX}}
\def \chan {{\it Chandra}}
\def \xmm {{\it XMM-Newton}}
\begin{frontmatter}



\title{Multi-fluid shocks in clusters of galaxies: entropy, $\sigma_{\rm v}$-$T$, $M$-$T$ and $L_{\rm X}$-$T$ scalings
}

\author[Ioffe]{A.M.Bykov}\ead{byk@astro.ioffe.ru},
\address[Ioffe]{A.F.Ioffe Institute for Physics and Technology, St.Petersburg, Russia, 194021}

\begin{abstract}
The nonthermal phenomena in clusters of galaxies are considered in
the context of the hierarchical model of cosmic structure
formation by accretion and merging of the dark matter (DM)
substructures. Accretion and merging processes produce large-scale
gas shocks. The plasma shocks are expected to be collisionless. In
the course of cluster's aggregation, the shocks, being the main
gas-heating agent, generate turbulent magnetic fields and
accelerate energetic particles via collisionless multi-fluid
plasma relaxation processes. The intracluster gas heating and
entropy production rate by a collisionless shock may differ
significantly from that in a single-fluid collisional shock.
Simple scaling relations for postshock ion temperature and entropy
as functions of shock velocity in  strong collisionless
multi-fluid shocks are presented. We show that the multi-fluid
nature of the collisionless shocks results in high gas
compression, reduced entropy production and modified $\sigma_{\rm
v}$-$T$, $M$-$T$ and $L_{\rm X}$-$T$ scalings. The scaling indexes
estimated for a simple model of a strong accretion multi-fluid
shock are generally consistent with observations. Soft X-ray and
extreme ultraviolet photons dominate the emission of strong
accretion shock precursors that appear as large-scale filaments.
Magnetic fields, turbulence and energetic particles constitute the
nonthermal components contributing into the pressure balance,
energy transport and emission of clusters.
Nonthermal emission of energetic particles could be a test to
constrain the cluster properties.

\end{abstract}

\end{frontmatter}

\section{Introduction}

Observational cosmology boosted by recent WMAP measurements of the
cosmic microwave background anisotropy, high-redshift supernovae
distribution, light element abundances  and impressive progress in
X-ray studies of clusters of galaxies is now entering   the stage
of the precision science (e.g. Ostriker \& Souradeep, 2004).
Clusters of galaxies being currently the largest gravitationally
bounded objects are used to study cosmic structure evolution and
to constrain the basic parameters of the concordance
($\Lambda$CDM) model (e.g. Bahcall, 2000). In the case of clusters
the precision cosmology approach requires an account of all the
physical components contributing to the energy density with
accuracy better than 10\%. Magnetic fields and energetic particles
in clusters contribute to the intracluster matter (ICM) pressure
at least at a level above a few percent and the actual
contribution could be substantially higher. Nonthermal radiation
of the clusters from radio to gamma rays and neutrino provide a
unique way to constrain physical parameters of the clusters.

Most of the cluster baryonic content is in the form of hot gas,
likely heated by a hierarchy of large scale shocks. The shocks
provide relaxation of accreting and merging matter flows in a course
of the hierarchical process of cluster formation. Collisionless
shock waves are the main plasma heating agent and also serve as an
universal source of energetic charged particles and radiation. We
shall discuss below some aspects of collisionless shocks of
plasma. Gas heating and entropy production in collisionless shocks
depends on both microscopic and macroscopic processes of shock
relaxation. Efficient nonthermal particle production results in
high gas compression and possibly efficient magnetic field
generation in the shock precursor but reduces postshock ion
temperature. Physical processes in the extended shock precursor
can be considered as a realization of gas preheating effects. That
effects may be reflected in the observable $\sigma_{\rm v}$-$T$, $M$-$T$
and $L_{\rm X}$-$T$ scalings.

\section{Collisionless shocks in clusters: magnetic fields, gas heating and entropy production}

Modern high resolution cluster observations in radio, IR, optical,
UV, and X-ray bands serve as probes to study the processes in the
ICM. Radio observations of electron synchrotron emission were the
first clear sign of the presence of GeV regime relativistic
electrons (positrons)  in clusters. The relativistic particles are
most probably related to the shock compression/acceleration
processes.

Energetic particles could be an essential component in clusters.
Nonthermal particle acceleration at shocks is expected to be an
efficient process at different evolutional stages of clusters.
Nonlinear wave-particle interactions being the governing process
of the SNR collisionless shock formation are responsible for both
shock heating and compression of the thermal gas as well as for
creation of energetic particle population.

A direct study of collisionless shock waves in a laboratory is an
extremely difficult task. Most of the experimental data on
collisionless shock physics are coming from space experiments.
There are direct observational data on the shock wave structure in
the interplanetary medium with clear evidences for ion and
electron acceleration by the shocks (e.g. Tsurutani and Lin, 1985).

Computer simulations of the full structure of collisionless shock
waves describe the kinetics of multi-species particle flows and
magneto-hydrodynamic (MHD) waves in the strongly-coupled system.
The problem is multi-scale. It requires a simultaneous treatment
of both "microscopic" structure of a subshock at the thermal
particle gyroradii scale where the injection process is thought to
occur, and an extended "macroscopic" shock precursor due to
energetic particles. The precursor scale is typically more than
$10^9$ times of the microscopic scale of the subshock transition
region.

\subsection{Micro and macro processes in multi-fluid shocks}

In the strong enough collisionless shocks (typically of a
Mach number above a few) resistivity cannot provide energy
dissipation fast enough to create a standard shock transition
(e.g. Kennel \etal, 1985) on a microscopic scale. Ion instabilities
are important in such shocks that are called supercritical.

At the microscopic scale the front of a supercritical shock wave
is a transition region occupied by magnetic field fluctuations of
an amplitude $\delta B/B \sim 1$ and characteristic frequencies of
about the ion gyro-frequency. Generation of the fluctuations is
due to instabilities in the interpenetrating multi-flow ion
movements. The width of the transition region of a
quasi-longitudinal shock wave reaches a few hundreds ion inertial
lengths defined as $l_i = c/\omega_{\rm pi} \approx 2.3 \times
10^7 n_{\rm a}^{-0.5}$ cm. Here $\omega_{\rm pi}$ is the ion
plasma frequency and $n_{\rm a}$ is the ionized ambient gas number
density measured in $\cmc$. The transition region of a
quasi-perpendicular shock is somewhat narrower. The wave
generation effects at the microscopic scale have been studied in
some detail by hybrid code simulations (e.g. Quest, 1988).
The large-amplitude magnetic field
fluctuation in the shock transition region were directly measured
in the interplanetary medium (see e.g. Kan \etal, 1990).

The reflected ions with a gyro-radius exceeding the width of the
shock transition region can be then efficiently accelerated, via
the Fermi mechanism, by converging plasma fluxes carrying MHD
fluctuations. The efficiency of the upstream plasma flow energy
conversion into nonthermal particles could be high enough
providing a hard spectrum of nonthermal particles up to some
maximal energy ${\varepsilon}_{\star}$. If the efficiency of ram
energy transfer to the energetic particles is high enough, an
extended shock precursor appears due to the incoming plasma flow
deceleration by the fast particle pressure. The precursor scale
$L$ is of the order of $(c/v_{\rm sh}) \cdot \lambda_{\star}$ --
orders of magnitude larger than the width of the shock transition
region. Here $\lambda_{\star}$ is the maximal mean free path of a
particle in the energy-containing part of the spectrum and $v_{\rm
sh}$ is the shock velocity. We shall later refer that energetic
particles as cosmic rays (CR).

The large scale ("macroscopic") structure of a collisionless shock
can be modeled by a two-fluid approach with a kinetic description
of nonthermal particles (see e.g. Blandford and Eichler 1987;
Malkov and Drury, 2001 for a review) or by a Monte Carlo method
(e.g. Jones and Ellison, 1991). In both methods some suitable
parameterization of particle scattering process must be postulated
{\it a priori}. It has been shown that the front of a strong
collisionless shock wave consists of an extended precursor and a
viscous velocity discontinuity (subshock) of a local Mach number
that is smaller than the total Mach number of the shock wave. The
compression of matter at the subshock can be much lower than the
total compression of the medium in the shock wave with allowance
for high compression in the precursor. We shall refer such shocks
later as CR-modified.

An important predicted feature of strong shocks with efficient CR
acceleration is a possibility to amplify an initial seed magnetic
field by orders of magnitude (e.g. Bell \& Lucek 2001; Bell 2004).
CR current upstream of the strong shock could drive magnetic
fluctuations on precursor scale length. The scale $L$ is  $\sim
(c/v_{\rm sh}) \cdot \lambda_{\star}$ which is expected to be
above a kpc, moreover the width $L \gsim$ 100 kpc for a cluster
scale shock. The pre-shock scale $L$ is $\gg$ 10$^9$ times longer
than the subshock transition region where strong magnetic field
fluctuations are directly produced by super-alfvenic bulk plasma
flows. The amplitude of the fluctuating magnetic field energy
density W$_{\rm B}$ is of the order of the shock accelerated CR
pressure which is in turn a substantial fraction of the shock ram
pressure $0.5\, \rho_1 \, v_{\rm sh}^2$. Here $\rho_1$ is the
shock upstream ambient gas density. For  typical cluster
parameters that provides a $\mu$G range magnetic field amplitude
in kilo-parsecs range scale CR-modified shock precursor. The
Faraday rotation measure $RM$ provided by a strong CR-dominated
shock in a cluster can reach values of $\gsim$ 10 rad m$^{-2}$ and
even higher. For the case of the so-called Bohm diffusion model
the measure $RM$ is proportional to the maximal energy of the ions
in the energy-containing part of the CR-spectrum accelerated by
the shock. Faraday rotation and synchrotron-Compton emission
measurements are used to estimate the magnetic fields in clusters
(e.g. Giovannini \& Feretti 2000; Carilly \& Taylor 2002; Newman
\etal, 2002).

Electron kinetics in collisionless shocks is different from that
of ions. Most of the observable emission comes from the electrons
so they require a careful study. Strong shocks  transfer the ram
kinetic energy of the flow into large amplitude nonlinear magnetic
fluctuations on a short scale of the transition region. The
thermal electron velocities in the ambient medium are higher than
the shock speed if the shock Mach number
 ${\cal M}_{\rm s}<\sqrt{m_p/m_e}$,
allowing for nearly-isotropic angular
distribution of the electrons. Non-resonant interactions of these
electrons with large-amplitude turbulent fluctuations in the shock
transition region could result in collisionless heating and
pre-acceleration of the electrons (Bykov and Uvarov, 1999). The
presence of large-amplitude waves in the shock transition region
 erodes many of the differences between quasi-parallel and
transverse shocks, providing the electron injection mechanism to
be similar for these shocks.

An exact modeling of a collisionless shock structure with an
account of the nonthermal particle acceleration effect requires a
nonperturbative self-consistent description of a multi-component
and multi-scale system including the strong MHD-turbulence
dynamics. That modeling is not feasible at the moment. Instead we
will use below some simplified description of a multi-fluid
strong shock structure with an appropriate parameterization of the
extended pre-shock and the gas subshock. Then some predicted
observable characteristics of the shocks can be confronted against
observational data. We will discuss first the effects of plasma
heating by modified shocks and then some specific predictions for
possible observational tests.

\subsection{Plasma ion heating by multi-fluid collisionless shocks in
clusters\label{sec:heat}}

The collisionless shock relaxation processes are multi-fluid. For
strong collisionless shocks in a magnetized plasma the nonthermal
particle acceleration effect is expected to be efficient and a
significant fraction of the ram pressure could be transferred into
high energy particles (ions only for the nonrelativistic
shocks). A strong shock of a total Mach number ${\cal M}_{\rm s}
\gg$ 1 consists of a broad precursor of a scale of $L \sim
(c/v_{\rm sh}) \cdot \lambda_{\star}$ due to  the upstream gas
deceleration by the nonthermal particle pressure gradient and a
viscous gas subshock of the modest Mach number ${\cal M}_{\rm sub}
\sim$ 3 (see for a review Blandford and Eichler, 1987; Jones and
Ellison, 1991; Malkov and Drury, 2001).

The total compression ratio $R_{\rm t}$ of a strong MHD shock
modified by an efficient nonthermal particle acceleration can be
estimated as
\begin{equation}
R_{\rm t} = \frac{\gamma + 1}{ \gamma - \sqrt{1 + 2(\gamma^2
-1)Q_{\rm esc}/\rho_1 v_{\rm sh}^3}}, \label{eq:crrt}
\end{equation}
assuming that the energy density in the shock upstream  is
dominated by the ram pressure and that the CR escape is through
the cut-off momentum regime (e.g. Malkov and Drury, 2001). Here
$Q_{\rm esc}$ is the energy flux carried away by escaping
nonthermal particles and $\gamma$ is the effective adiabatic
exponent.

 The distribution function of nonthermal particles and
the bulk flow profile in the shock upstream region are sensitive
to the total compression ratio $R_{\rm t}$. Thus, the exact
calculation of the escape flux $Q_{\rm esc}$ can be performed only
in fully nonlinear kinetic simulations. Nevertheless, some
approximate iterative approach can be used to make the
distribution function consistent with the shock compression. The
subshock is the standard gas viscous shock of a Mach number ${\cal
M}_{\rm sub}$. For that simplified steady-state model of a strong
CR-modified shock the downstream ion temperature $T^{(2)}_{\rm i}$
can be estimated (Bykov, 2004) for the shock of a given velocity 
if $R_{\rm t}$ and ${\cal M}_{\rm sub}$ are known:
\begin{equation}
T^{(2)}_{\rm i} \approx \phi({\cal M}_{\rm sub})  \cdot \frac{
\mu~ v_{\rm sh}^2}{\gamma_{\rm g}R_{\rm t}^2(v_{\rm sh})} ,~~{\rm
where}~~ \phi({\cal M}_{\rm sub}) = \frac{2 \gamma_{\rm g} {\cal
M}_{\rm sub}^2 - (\gamma_{\rm g} -1)}{(\gamma_{\rm g} -1){\cal
M}_{\rm sub}^2 + 2}. \label{eq:tcr}
\end{equation}

Single fluid strong shock heating represents the limit ${\cal
M}_{\rm sub} = {\cal M}_{\rm s} \gg$ 1, since there is no
precursor in that case. In a single fluid strong shock with
${\cal M}_{\rm s} \gg 1$ and  ${\cal M}_{\rm a} \gg 1$ we have
\begin{equation}
 T^{(2)}_{\rm i} \approx 2\cdot \frac{(\gamma_{\rm g}
-1)}{(\gamma_{\rm g} +1)^2}~\mu v_{\rm sh}^2 = 1.38 \cdot
10^7~v^2_{s8}~(K), \label{eq:rht1}
\end{equation}
 for any magnetic field
inclination (e.g. Draine and McKee, 1993).  The $\gamma_{\rm g}$ is
the gas adiabatic exponent, the mass per particle $\mu$ was
assumed to be $[1.4/2.3] \cdot m_H$ and $v_{\rm s8}$ is the shock
velocity in 10$^8~\cmsec$. In the single-fluid system the
compression ratio $R_{\rm t} = R_{\rm sub} \rightarrow
(\gamma_{\rm g} +1)/(\gamma_{\rm g} -1)$ does not depend on the
shock velocity and the Eq.(\ref{eq:tcr}) reduces to
Eq.(\ref{eq:rht1}). However in the multi-fluid shocks the total
compression ratio depends on the shock velocity and could be
higher than that in the single-fluid case. That implies somewhat
lower postshock ion temperatures for the strong multi-fluid shock
of the same velocity and could be tested observationally (e.g. Bykov, 2004).

Consider multi-fluid CR-modified strong shocks. The gas heating
mechanism in a CR-modified shock precursor must be specified in
a simplified two-fluid approach, providing a connection
between the total compression $R_{\rm t}$ and the gas subshock
Mach number ${\cal M}_{\rm sub}$. It is worth to recall the plasma
parameter $\beta = {\cal M}^2_{\rm a}/{\cal M}^2_{\rm s}$ is also
the ratio of the thermal and magnetic pressures.

The gas heating mechanism in the modified shock precursor is still
under debate (e.g. Malkov and Drury, 2001) and is usually
postulated in shock modeling. There are extensive simulations of
the models with {\it adiabatic heating} of the shock precursor
(e.g. Kang \etal, 2002). The {\it \alf waves dissipation}
mechanism was suggested by McKenzie and V\"olk (1982). The
turbulent gas heating due to {\it acoustic instability wave
dissipation} could be more efficient, however, if the acoustic
instability develops (e.g. Malkov and Drury, 2001).

The subshock has ${\cal M}_{\rm sub} \gsim 3$ in the most of the
simulated strong shocks with adiabatic heating of the precursor.
The result is not too sensitive to the ion injection rate if the
rate exceeds some threshold to provide the shock modification by
the accelerated ions. The scaling ${\cal M}_{\rm sub} \sim 2.9
\cdot {\cal M}_{\rm s}^{0.13}$ was obtained by Kang \etal, (2002).
They also found that $R_{\rm t} \propto {\cal M}_s^{3/4}$ is a
reasonably good approximation for the two-fluid time-dependent
simulations of a CR-modified shock with adiabatic precursor.
However, if  precursor heating due to \alf waves dissipation or
the acoustic wave instability is efficient then for the plasma
compression in the shock precursor could become less efficient.

Notice that the approximations for the shock compression and
heating given above were obtained within a simplified two-fluid
steady-state approach where the Bohm-like diffusion model was
postulated.  The possibility of a shock acoustic instability
development is still an open question. It requires a dedicated
study of the connection between the diffusion coefficient and the
local flow parameters. Recent detailed numerical studies of the
unsteady CR-modified shocks by Kang \etal, (2002) {\it a priori}
assumed the spatial dependence of the diffusion coefficient
$k(p,x) \propto n^{-1}(x)$ to prevent the acoustic instability.
Gas heating due to acoustic instability waves or \alf waves
dissipation would change their results.

A modest heating rate is expected in the case of adiabatic plasma
compression in the precursor. That could be the case for very high
\alf Mach number shocks $1 \ll {\cal M}_s^2 \ll {\cal M}_{\rm a}$
(i.e. $1\ll {\cal M}{\rm _s} \ll \sqrt\beta$) where the \alf wave
dissipation is not efficient and the acoustic wave instability
does not develop. The total compression ratio is $R_{\rm t} \sim
{\cal M}_{\rm s}^{3/4}$ (e.g. Blandford and Eichler, 1987). Then
the downstream ion temperature for the case of a weak upstream
magnetic field (and ${\cal M}_{\rm sub} \sim 3$) can be estimated
from:
\begin{equation}
T^{(2)}/ T^{(1)} \approx 2.2 \cdot \sqrt{\cal M}_{\rm s} (1 +
f_{\rm ei})^{-1}. \label{eq:tia2}
\end{equation}
Here $f_{\rm ei} = T_{\rm e}/T_{\rm i}$. The latter case of ${\cal
M}_{\rm s} \ll \sqrt\beta$ would require the plasma parameter
$\beta > 100$ in the shock far upstream. The scaling can be also
relevant to the internal shocks of ${\cal M}_{\rm s} \lsim 10$.
Following cosmological shocks simulations of Ryu \etal, (2003) the
internal shocks are expected to be rather numerous in the ICM.
Note here that the same temperature scaling $T^{(2)}/ T^{(1)}
\propto \sqrt{\cal M}_{\rm s}$ is appropriate for strong radiative
shocks (see e.g. Bouquet \etal, 2000).

Under an assumption that the main heating mechanism of the gas in
the precursor region is due to \alf waves dissipation, an
asymptotical estimation for the total compression of the shock
$R_{\rm t} \approx 1.5 {\cal M}_{\rm a}^{3/8}$ was presented by
Berezhko and Ellison (1999). The estimation was obtained assuming
a constant \alf speed in the shock upstream and it is valid if
${\cal M}_{\rm s}^2 \gg {\cal M}_{\rm a} \gg 1$ in the far
upstream flow. One may then obtain the ion temperature just behind
the strong modified shock with {\it the preshock heating due to
\alf wave dissipation}~(Bykov, 2002) from Eq.(\ref{eq:tcr}):
\begin{equation}
T^{(2)}_{\rm 6} \approx 0.32 \cdot \phi({\cal M}_{\rm sub}) \cdot
v_{\rm s8}^{5/4} n_{\rm a}^{-3/8} B_{-6}^{3/4}(1 + f_{\rm
ei})^{-1}. \label{eq:tia1}
\end{equation}
The ion temperature is measured in 10$^6$ K. Eq.(\ref{eq:tia1}) is
valid under the conditions of ${\cal M}_{\rm s} \gg 1 $ and ${\cal
M}_{\rm s} \gg \sqrt {\beta} $ in the far upstream flow. We have
$0.32 \cdot \phi(3) \approx 1.2$ for $\gamma_{\rm g}$= 5/3.

An important distinctive feature of the strong CR-modified shock
with gas heating by \alf wave dissipation (given by
Eq.(\ref{eq:tia1})) is the dependence of the postshock temperature
on the number density and magnetic field of the incoming plasma
$\propto (B/\sqrt n)^{3/4}$. There is a remarkable difference
between the single-fluid shock heating where $T^{(2)} \propto
v_{\rm sh}^2$ for ${\cal M}_s^2 \gg 1$ and that for the
CR-dominated multi-fluid strong shocks (given by
Eq.(\ref{eq:tia1}) and Eq.(\ref{eq:tia2})) where $T^{(2)} \propto
v_{\rm sh}^a$.

It is convenient to introduce the scaling $R_{\rm t}(v_{\rm sh})
\propto v_{\rm sh}^{\Delta}$ to describe the different cases of
strong shock heating. Then $T_i^{(2)} \propto \phi({\cal M}_{\rm
sub})\cdot v_{\rm sh}^{2(1-\Delta)}$ and the index $a =
2(1-\Delta) + \delta$. Here the index $\delta$ approximates the
velocity dependence of $\phi({\cal M}_{\rm sub}) \propto v_{\rm
sh}^{\delta}$. The subshock Mach number ${\cal M}_{\rm sub}$
depends, in general, on ${\cal M}_{\rm s}$ and ${\cal M}_{\rm a}$.
For the \alf wave heating case (see Eq.(\ref{eq:tia1})) we have
$\Delta \approx 3/8$, the index $\delta$ is typically $\leq 0.25$,
and thus $a \leq 1.5$. In the case of high enough efficiency of
the turbulent heating of gas due to the acoustic instability of a
strong shock the index $a$ may exceed 1.5 being closer to the
single-fluid limit. The shock velocity dependence of $R_{\rm
t}(v_{\rm sh})$ and $T^{(2)}_i(v_{\rm sh})$ discussed above should
result in cluster thermodynamic parameter scalings $\sigma_{\rm
v}$-$T$ and other to be confronted against observational data (see
e.g. Rosati, Borgani \& Norman 2002 for a review). The scaling
relations depend on electron-ion relaxation and we will discuss
them below.

\subsection{ICM entropy production by multifluid accretion shocks }

Cold gas falling into the DM dominated gravitational well passes
through a strong accretion shock. The shock is a source of gas
entropy production in the ICM (e.g. Knight \& Ponman, 1997; Tozzi
\& Norman, 2001; Voit \etal, 2003). The post-shock entropy $K =
K_0\, T/\rho^{2/3}$ used in the ICM analysis and simulations (e.g.
Bialek \etal, 2001) is related to the standard thermodynamic
entropy $s$ through $K \propto \exp(s/c_v)$. In the standard
scenario with a single-fluid accretion shock the post-shock
entropy  $K_{\rm sf} \propto v_{\rm sh}^2 \cdot \rho_1^{-2/3}$
(e.g. Voit \etal, 2003).

The multi-fluid nature of the collisionless accretion shock modifies
the standard scaling relation to be
\begin{equation}
K_{\rm mf} \propto v_{\rm sh}^2\cdot [R_{\rm t}(v_{\rm sh})]^{-(1 +\gamma_{\rm
g})}\cdot \phi({\cal M}_{\rm sub})\, \cdot \rho_1^{(1-\gamma_{\rm
g})}
\end{equation}
The compression ratio in CR-shocks is higher than in a strong
single-fluid shock of the same velocity resulting in reduced
post-shock entropy production. For example, in the case of
\alf heating the post-shock entropy of a multi-fluid shock reduces
as $K_{\rm mf}/K_{\rm sf} \sim (15/{\cal M}_{\rm a})$ for ${\cal
M}_{\rm a} > 15$ and ${\cal M}_{\rm s}^2 > {\cal M}_{\rm a}$. Here
and below in numerical estimations we assume $\gamma_{\rm g}=5/3$
though non-thermal baryonic component could reduce the index
$\gamma_{\rm g}$.

 Since $R_{\rm t}(v_{\rm sh})$ and $\phi({\cal M}_{\rm sub})$
 are shock velocity
dependent the simple scaling $K \propto v_{\rm sh}^2 \cdot
\rho_1^{-2/3}$ is not valid.  In CR-modified shocks $K_{\rm mf}
\propto v_{\rm sh}^{\nu} \cdot \rho_1^{(1-\gamma_{\rm g})}$ or
$K_{\rm mf} \propto T^{\nu/a}$, where $\nu =2 - (1+\gamma_{\rm
g})\cdot \Delta +\delta$. For the case of  \alf wave heating the
index $\nu \lsim 1.25$ and $K_{\rm mf} \propto T^{0.8}$ assuming
$\gamma_{\rm g} = 5/3$. Recently Piffaretti \etal, (2004) found
that the dispersion in the observed cluster entropy profiles is
smaller if an empirical relation $K \propto T^{0.65}$ is used
instead of the standard $K \propto T$.

Consider a simple model of smooth accretion of cold gas
 through a strong accretion shock by Voit \etal, (2003).
 The gas of velocity $v_{\rm ac}$ accretes at a rate ${\dot M}_g$ through the
 shock at a radius $r_{\rm ac}$ where
\begin{equation}
{\dot M}_g = 4\pi r_{\rm ac}^2 \rho_1 v_{\rm ac},~~~ v_{\rm ac}^2=2GM\xi
r_{\rm ac}^{-1},~~~\xi = 1 - r_{\rm ac}/r_{\rm ta}. \label{eq:acr1}
\end{equation}
Here $M(t)$ is the cluster mass and $r_{\rm ta}$ is the matter
turnaround radius. Then the entropy $K_{\rm mf}$ just behind the
multi-fluid shock is expressed through $T^{(2)}_i(v_{\rm ac})$ and
$\rho_2 = R_{\rm t}(v_{\rm ac})\cdot \rho_1$. In the \alf wave
heating case $K_{\rm mf}(t)\propto (Mt)^{(1+\delta)/3}$, instead
of $K_{\rm sf}(t) \propto (Mt)^{2/3}$ in the single-fluid regime.
Multi-fluid shock results in a slower post-shock entropy
production. As we have noted above the regime of CR-shock
compression depends on the plasma parameter $\beta$ in the
infalling gas. The plasma parameter $\beta$ is currently poorly
known because the intercluster magnetic fields are not well
constrained. The effects of shock modifications are important for
both the models of smooth accretion of cold gas and for accretion
of hierarchical structures as well.

Preheating of accreting gas by different physical processes (e.g.
due to early star formation in protocluster region) was suggested
by Evrard \& Henry (1991), as a possible reason for the breaking
of the scaling relations for pure gravitational cluster
compression (Kaiser, 1986). The observed high metallicity of
clusters at different  redshifts indicates that strong starburst
activity was highly likely at some stage. The preheating produces
some initial level of gas entropy ("entropy floor") (see e.g.
extensive simulations by Bialek \etal, 2001; Borgani \etal, 2001).
Multi-fluid strong shocks provide another natural way of
preheating of accreting gas.

We have discussed above the ion temperature and entropy
production in multi-fluid shocks. However, cluster radiation
spectra (X-ray and multi-wavelength) depend in large on the
electron temperature.

\subsection{Electron temperature in the multifluid shocks }

The electron and ion temperatures eventually equilibrate  in a
thick postshock layer due to Coulomb collisions. The complete
$e-i$ Coulomb equilibration requires the system age $t \gsim
10^{10}~T_6^{3/2}/n$, where the postshock density $n$ is measured
in $\cmc$, the ion temperature $T_6$ is in 10$^6$ K, and $t$ is in
seconds (see e.g. Mewe 1990 for a review). The Coulomb
equilibration rate is fast enough to guarantee $T_e \approx T_i$
on the cluster scale for a Hubble time.

The complete $e-i$ equilibration is not always the case for a
local vicinity of a collisionless shock, however. For the internal
shocks in the ICM the electron-ion temperature relaxation layer
can be of the order of 100-300 kpc and it can be even larger for
the external shocks. It is highly likely that the $e-i$ Coulomb
equilibration will be complete in the extended precursors of
multi-fluid accretion shocks.  \alf waves and magnetic fluctuation
pre-heat the ions to temperatures of about 0.1--0.5 keV in the
extended shock precursor. The column depth in the precursor could
be about $\gsim 10^{19} \cms$ providing substantial $e-i$
equilibration in the gas entering the subshock of ${\cal M}_{\rm
sub} \sim 3$. Then ions will be heated in the collisionless
subshock by a factor of $\sim$ 3.5.

The gas density and ion temperature jumps do not necessarily
coincide with that of the electron temperature. X-ray telescopes
can resolve in principle the extended structure with gradual
electron temperature profile behind shock front. On the other hand
in case of merger events it is more easy to identify the "cold
fronts" -- sharp contact discontinuities between cool gas bodies
moving through a hot ICM. Such structures were discovered in A2142
(Markevitch \etal, 2000), A3667 (Vikhlinin \etal, 2001) and other
clusters.

It should be noted, however, that while the Coulomb collisions are
unavoidable they are not the only possible electron heating
process in shocks. Electromagnetic field fluctuations providing
the ion flow relaxation in collisionless shock may heat electrons
on rather a short scale.

The initial electron temperature just in the shock downstream
depends on  collisionless electron heating (see e.g. Cargill and
Papadopulos, 1988; Bykov and Uvarov, 1999). Non-resonant
interactions of the electrons with strong nonlinear fluctuations
generated by kinetic instabilities of the ions in the transition
region inside the shock front may play the main role in the
heating and pre-acceleration of the electrons, as it was shown in
the model by Bykov and Uvarov (1999). They calculated the electron
energy spectrum in the vicinity of  shock waves and showed that
the heating and pre-acceleration of the electrons occur on a scale
of the order of several hundred ion inertial lengths $l_i$ in the
vicinity of a viscous discontinuity. Although the electron
distribution function is significantly non-equilibrium near the
shock front, its low energy part can be approximated by a
Maxwellian distribution. The effective electron temperature just
behind the front, obtained in this manner, increases with the
shock wave velocity as $T_{\rm e} \propto v_{\rm sh}^b$ with $b
\leq 2$. They also showed that if the electron transport in the
shock transition region is due to turbulent advection by strong
vortex fluctuations on the ion inertial length scale ($l_i$), then
the nonresonant electron heating is rather slow (i.e. $b \leq
0.5$), but the electrons are still injecting into the diffusive
Fermi acceleration. A highly developed vortex-type turbulence is
expected to exist in the transition regions of strong shocks. That
would imply that the initial $T_{\rm e}/T_{\rm i} \propto v_{\rm
sh}^{(b-a)}$ just behind the transition region would decrease with
the shock velocity for ${\cal M}_{\rm s} \gg 1$. So, the electron
heating is more efficient in the shocks of lower velocities. There
are some observational indications that $T_{\rm e}/T_{\rm i}
\propto v_{\rm sh}^{-1}$ for collisionless shocks in supernova
remnants (e.g. Raymond, 2001). The scaling is generally consistent
with the interplanetary shock data compiled by Schwartz \etal,
(1988) (see also Rakowski \etal, 2003).
The degree of electron-ion equilibration in collisionless
shock is found to be a declining function of shock speed. In the
case of strong vortex-type turbulence in the shock transition
region $(a - b) \sim $ 1 for the \alf heating case described
above.

\section{Effect of the non-thermal components on $\sigma_{\rm v}$-$T$, $M$-$T$, $L_{\rm X}$-$T$ relations}

X-ray cluster studies accompanied by multi-wavelength observations
have provided a possibility to investigate statistical and
evolutional properties of ICM in clusters (see e.g. Forman \&
Jones, 1982; Sarazin, 1988; Rosati, Borgani \& Norman, 2002 for a
review). In particular, relationships between the optical measured
galaxy velocity dispersion $\sigma_{\rm v}$, the X-ray bolometric
luminosity $L_{\rm X}$, the X-ray temperature $T_X$ and cluster
gravitating mass $M$ were determined observationally. Simulated
scalings are to be confronted against observational data (see e.g.
Arnaud \& Evrard 1999, Mohr, Mathiesen \& Evrard, 1999, Wu \etal,
1999, and Rosati, Borgani \& Norman, 2002 for a review) to
constrain the models of cosmic structure evolution and to estimate
the baryon fraction.

The $L_{\rm X}$-$T$ and other relations were simulated by
Cavaliere, Menci \& Tozzi (1998, 1999) using the standard
single-fluid Rankine-Hugoniot law. They considered hierarchical
merging history using the generalized Press-Schechter DM halo
distribution. That allows the authors to make predictions for
the expected cosmological evolution of the $L_{\rm X}$-$T$ correlation. We
will discuss here some simplified models to study the effect of
multi-fluid shocks on the thermodynamic relations. The shock velocity
dependences of the compression ratio $R_{\rm
t}(v_{\rm sh})$ and post-shock temperature $T^{(2)}_i(v_{\rm sh})$
discussed above result in modified cluster thermodynamic parameter
scalings $\sigma_{\rm v}$-$T$, $M$-$T$, $L_{\rm X}$-$T$.

In a simplified model of smooth accretion of cold gas by Voit
\etal\ (2003), a strong accretion shock of a velocity given by
Eq.(\ref{eq:acr1}) will have $v_{\rm sh} \propto \sigma_{\rm v}$.
Lumpy accretion of DM subhalos produce internal (merger) shocks
inside the ICM gas heated by the external (accretion) shock. The
properties of shocks in the large-scale structure of the Universe
were simulated in the context of the $\Lambda CDM$-cosmology using
PM/Eulerian hydrodynamic codes (e.g. Miniati \etal, 2000; Ryu
\etal, 2003). Internal shocks with $2 \leq {\cal M}_{\rm s} \leq
4$ were found to be most important in energy dissipation providing
ICM heating. The code deals with N-body CDM and  single-fluid
gas dynamics. However, if a strong accretion shock is multi-fluid,
providing reduced post-shock ion temperature and entropy, then the
internal shocks could have systematically higher Mach numbers $4
\leq {\cal M}_{\rm s} \leq 8$.

Consider as a generic example accretion of a cold gas of
$n_{\rm a} \sim 10^{-5} \cmc$ and $B \sim 0.03-0.1~\mu G$ through
an accretion shock located at $r_{\rm ac} \sim$ 2 Mpc around a
rich cluster of $M \sim 10^{15} \Msun$. The gas velocity in the
shock rest frame is $v_{\rm sh} \sim 2,000 \kms$. Then the gas
compression ratio by the strong multi-fluid shock $R_{\rm t} \sim
8$ and the post-shock ion temperature $T^{(2)}_{\rm i} \sim$ keV
can be obtained from Eq.(\ref{eq:tia1}). Magnetic field can be amplified
in the
shock precursor and compressed by a sub-shock providing $B \sim
\mu G$ in the post-shock flow. The extended multi-fluid shock
precursor of a scale of a few Mpcs and of a width of some hundred
kpcs will have a temperature of about 0.2--0.3 keV and magnetic
field $\sim$ 0.5 $\mu G$.

Further heating and compression of the accreted gas in the inner
cluster will be provided by internal shocks. Though an internal shock
of ${\cal M}_{\rm s} \sim 6$ is not extremely strong, it may still
be multi-fluid. For such a shock Eq.(\ref{eq:tia1}) is valid under
the conditions of ${\cal M}_{\rm s} \gg 1 $ and ${\cal M}_{\rm
s}^2 \gg {\cal M}_{\rm a}$ in the far upstream flow of the inner
shock. The later condition can be expressed as $v_{\rm s8} \gg
0.6~n_{\rm -4}^{1/2}\;T_{\rm 7}\; B_{\rm -6}^{-1}$. The ion
temperature $T_{\rm 7}$ is measured in 10$^7$ K. Then considering
the shocks as the main heating agents for the inner ICM one can
estimate thermodynamic scaling relations.

\subsection{$\sigma_{\rm v}$-$T$ scaling}
For a cluster near the virial equilibrium the internal shock
velocity $v_{\rm sh} \propto \sigma_{\rm v}$.
 The multi-fluid nature of CR-modified shocks would result in
$\sigma_{\rm v} \propto T_X^{1/a}$ for the cluster with relaxed
$e-i$ temperatures. In the case of the shock with \alf wave
dissipation heating $a \leq$ 1.5, and  the $\sigma_{\rm v}$-$T$
scaling index is $\gsim 2/3$. The statistical analysis of a sample of
about 100 observed clusters given by Wu \etal\ (1999) provided the
scaling $\sigma_{\rm v} = 10^{2.49 \pm 0.03}\;T^{0.64\pm 0.02}$
consistent with $\sigma_{\rm v} \propto T_X^{2/3}$.

\subsection{$M$-$T$ scaling}

For a virialized cluster of a total mass $M$ we have $\sigma_{\rm
v}^2 \propto M^{2/3}$. Therefore, in the model of a multi-fluid 
shock $M \propto T^{3/a}$ is expected. For the case of relaxed
$e-i$ temperatures (on the cluster scale) that results in $M_{\rm
ICM} \propto T_X^{3/a} \cdot f_{\rm b}$. Here the ICM gas mass
$M_{\rm ICM} = f_{\rm b} \cdot M$, and $f_{\rm b}$ is the baryon
gas mass fraction. The index of $M_{\rm ICM}$-$T$ scaling derived
by Mohr \etal\ (1999) from observations is 1.98 $\pm$ 0.18,
consistent with the  value $\geq$ 2 predicted by the \alf wave
heating multi-fluid shock model ($a \leq 1.5$). The $M$-$T$
relation for a sample of galaxy clusters, groups and elliptical
galaxies was studied by Sanderson \etal\ (2003). The authors
reported the $M$-$T$ index 1.84 $\pm$ 0.06 for the sample. To
apply our simple model to a sample of the objects of different
cosmological ages one has to account for the evolution of the mean
density of the Universe with redshift and for galaxy feedback
processes (see e.g. Borgani \etal, 2004).

\subsection{$L_{\rm X}$-$T$ scaling}
The model $L_{\rm X}$-$T$ scaling can be estimated from the
relation $L_{\rm X} \propto \sigma_{\rm v}^4 \cdot T^{1/2}\cdot
f_{\rm b}^2\cdot r_{\rm Xc}^{-1}$ given by Wu \etal\ (1999). Here
$r_{\rm Xc}$ is the cluster core radius in spherical isothermal
$\beta$-model of ICM gas distribution (of index 2/3). Thus, in the
multi-fluid shock model we have $L_{\rm X} \propto T^{4/a + 1/2}$.
For the preshock heating by \alf waves $a \leq 3/2$ and the
$L_{\rm X}$-$T$ index is $\geq 3.16$. We did not account here for
the weak temperature dependence of the thermal bremsstrahlung
Gaunt-factor. Arnaud \& Evrard (1999) obtained the $L_{\rm X}$-$T$
index 2.88 $\pm$ 0.15 from a sample of 26 clusters, while
Markevitch (1998) obtained 2.64 $\pm$ 0.27 removing photons from
central regions of the clusters. It should be noted that contrary
to the $\sigma_{\rm v}$-$T$ and $M_{\rm ICM}$-$T$ relations the
$L_{\rm X}$-$T$ relation is more sensitive to the assumed
spherical symmetry and  matter distribution inside the cluster.

\section*{Summary}

The multi-fluid processes of shock heating discussed above are a
natural realization of the idea of accreting gas preheating.
Collisionless shocks generate energetic non-thermal particles that
can penetrate far into the upstream flow. The particles decelerate
the flow and preheat the gas. They can also efficiently generate
fluctuating magnetic fields of $\sim \mu$G range on some
kilo-parsecs scale.

An important distinctive feature of multi-fluid shocks is their
high gas compression $R_{\rm t}(v_{\rm sh})$ that could be well
above the single fluid shock limit $(\gamma_{\rm g}
+1)/(\gamma_{\rm g}-1)$. At the same time entropy production for a
strong multi-fluid shock scales as $R_{\rm t}(v_{\rm
sh})^{-(\gamma_{\rm g} +1)}$ and it is significantly reduced
compared to the single-fluid shock of the same velocity. The
effects are due to energetic particle acceleration and magnetic
field generation. Energetic particles can penetrate into the shock upstream
region. They are coupled with the upstream gas through fluctuating
magnetic fields (including the \alf waves). Magnetic field
dissipation  provides gas preheating and entropy production in the
shock precursor. Energetic particles evacuate gas energy and
momentum resulting in high gas compression and reduced
temperature. Such a heated pre-shock region would appear as an
extended filament of a width $L \sim (c/v_{\rm sh}) \cdot
\lambda_{\star} \gsim 3\times 10^{14} \cdot \epsilon_{\star} \cdot
B_{-6}^{-1}$ cm. Here $\epsilon_{\star}$ (in GeV) is the highest
energy of the hard branch of the accelerated particle spectrum.
Since $B_{-6} \lsim$ 1 in the cluster outskirts and if the hard
spectrum of energetic nuclei extends till $\sim 10^9$ GeV (c.f.
Norman, Melrose \& Achterberg, 1995) we have $L \gsim$ 100 kpc and
even wider. A warm gas ($\sim$ 0.2 keV) emission filament found
with \xmm\ in the outskirts of Coma cluster by Finoguenov \etal\
(2003) could be an extended heated precursor of a strong
multi-fluid accretion shock.

Energetic nuclei can be stored in cluster magnetic fields for
Hubble times (e.g. Berezinsky, Blasi \& Ptuskin, 1997). Nonthermal
particles and magnetic fields contribute to the total pressure of
the ICM. Nonthermal electromagnetic emission and neutrinos due to
energetic particles interactions potentially provide tests to
constrain the ICM nonthermal components. Diffuse synchrotron radio
halos were reliably detected in some clusters (e.g. Giovannini \&
Feretti, 2000; Carilli \& Taylor, 2002) indicating the presence of
highly relativistic electrons. There are some evidences for a
presence of excesses above the dominating thermal emission of the
ICM hot gas in both hard X-rays and EUV. Brunetti \etal\ (2004) argued
that present observations of non-thermal radiation from clusters of galaxies
can be explained even if the pressure of relativistic hadrons in
the intracluster medium does not exceed 5--10$\%$ of that of the thermal gas.
A reliable detection or
a meaningful constrain of the hard X-ray tails with {\it INTEGRAL}
or {\it ASTRO-E2} and future gamma-ray observations with {\it GLAST}
would be very important to identify the nature
of the nonthermal processes in clusters. The existing upper limits
for emission above 100 MeV obtained with {\it CGRO EGRET} (e.g.
Reimer \etal, 2003) can not yet rule out  substantial energy density
of energetic nuclei (see also Blasi, 1999).
A population of energetic nuclei of energy
$\lsim$ GeV is hard to detect with current telescopes even if they
have a substantial pressure comparable with that of the hot thermal
gas. Ultra high energy CR acceleration above 10$^9$ GeV by cosmic
shocks was suggested by Norman, Melrose \& Achterberg (1995) as a
plausible scenario.

Study of the cooling flows spectra with {\it RGS} and spatially
resolved X-ray spectroscopy with {\it EPIC} aboard \xmm\  (e.g.
Kaastra \etal, 2004) revealed significant systematic deficit of the
emission of the low temperature gas as compared to the isobaric
cooling flow model predictions. Magnetic fields should be
progressively more important in the cluster central region. While
the strong multi-fluid shocks are expected to dominate at the
outskirts of the cluster the hierarchical clustering model
predicts  moderate and weak shocks in the inner parts of the
cluster as it was demonstrated by Ryu \etal, (2003). One may also
consider a magnetic field reconnection induced by cluster
substructure motions. In the scenario the local energy release
rate due to magnetic fields reconnection
 would be $\propto B^{c} \propto n_a^c$ with $c \geq$ 2.
Dissipation of the weak shocks from the reconnection events
and that produced by
fast moving substructures
 would preferentially
produce energetic particles thus allowing for the non-local
heating effect different from the standard heat-conduction  (Bykov, 2002).
The cold-phase filling factor could be strongly reduced in that case.

\section*{Acknowledgements}
I thank the referees for constructive comments. I am grateful to
{\it COSPAR} for travel support. The work was partially supported
by RBRF 03-02-17433 and 04-02-16595 grants.

\section*{References}

\begin{list}{}{}

\item Arnaud, M. and Evrard, A.E. The $L_{\rm X}-T$ relation and
intracluster gas fractions of X-ray clusters, {\em MNRAS}, 305,
631-640, 1999.
\medskip
\item
Bahcall. N.,  Clusters and cosmology, {\em Physics Reports},
333/334, 233-244, 2000.
\medskip
\item
Bell, A. R., Lucek, S. G. Cosmic ray acceleration to very high
energy through the non-linear amplification by cosmic rays of the
seed magnetic field, {\em MNRAS}, 321, 433-438, 2001.
\medskip

\item
Bell, A. R., Turbulent amplification of magnetic field
and diffusive shock acceleration of cosmic rays, {\em MNRAS}, 353,
550-558, 2004.
\medskip

\item
Berezinsky, V.S., Blasi, P., Ptuskin, V.S. Clusters of galaxies as
storage room for cosmic rays, {\em Astrophys. J.}, 487, 529-535,
1997.
\medskip

\item
Berezhko, E.G., and Ellison, D.C.  A simple model of nonlinear
diffusive shock acceleration, {\em Astrophys. J.}, 526, 385-399,
1999.
\medskip

\item
Bialek, J.J., Evrard, A.E., Mohr,J.J.,
 Effects of preheating on X-Ray scaling relations in galaxy clusters,
{\em Astrophys. J.}, 555, 597-612, 2001.
\medskip

\item
Blandford, R., and D.Eichler,  Particle acceleration at
astrophysical shocks: A theory of cosmic ray origin, {\em Phys.
Rep.}, 154, 1-75, 1987.
\medskip

\item
Blasi, P., On the equipartition of thermal and nonthermal energy in clusters of galaxies, {\em Astrophys. J.}, 525, 603-608, 1999.
\medskip

\item
Borgani, S., Governato, F., Wadsley, J., Menci, N., Tozzi, P.,
Lake, G., Quinn, T., Stadel, J., Preheating the intracluster
medium in high-resolution simulations: The effect on the gas
entropy , {\em Astrophys. J.}, 559, L71-L74, 2001.
\medskip

\item
Borgani, S., Murante, G., Springel, V., Diaferio, A.,
Dolag, K., Moscardini, L., Tormen, G., Tornatore, L.,
Tozzi, P., X-ray properties of galaxy clusters and groups
from a cosmological hydrodynamical simulation,
{\em MNRAS}, 348, 1078-1096, 2004.
\medskip

\item
Bouquet, S., R, Teyssier, and J.P. Chieze, Analytical study and
structure of a stationary radiative shock, {\em Astrophys. J.
Suppl.}, 127, 245-252, 2000.
\medskip

\item
Brunetti, G., Blasi, P., Cassano, R., Gabici, S.,  Alfvenic reacceleration of relativistic particles in galaxy clusters: MHD waves, leptons and hadrons,
{\em MNRAS}, 350, 1174-1194, 2004.
\medskip

\item
Bykov, A.M.,  Uvarov, Yu.A.,  Electron kinetics in collisionless
shock waves, {\em JETP}, 88, 465-475, 1999.
\medskip

\item
Bykov, A.M., Nonthermal processes in clusters of galaxies, in:
{\em The Gamma-Ray Universe} eds. A.Goldwurm, D.N.Neumann,J.Tran
Thanh Van, The GIOI Publishers, 341-350, 2002.
\medskip

\item
Bykov, A.M., Shocks and particle acceleration in SNRs: theoretical aspects,
{\em Adv. Space Res.}, 33, 366-375, 2004.
\medskip

\item
Cargill, P.\,J., K. Papadopoulos,  A mechanism for strong
shock electron heating in supernova remnants, {\em Astrophys. J.},
329, L29-L32, 1988.
\medskip

\item Carilli, C. L. and Taylor, G. B.,
Cluster Magnetic Fields,
{\em Ann. Rev. Astron. Astroph.}, 40, 319-348, 2002.
\medskip

\item Cavaliere, A., Menci, N. and
Tozzi, P. Diffuse baryons in groups and clusters of galaxies, {\em
Astrophys. J.}, 501, 493-508, 1998.
\medskip

\item Cavaliere, A., Menci, N. and
Tozzi, P. Hot gas in clusters of galaxies: the punctuated
equilibria model, {\em MNRAS}, 308, 599-608, 1999.
\medskip

\item
Draine, B.T. and C.F. McKee,  Theory of interstellar shocks, {\em
Ann. Rev. Astron. Astroph.}, 31, 373-432, 1993.
\medskip

\item Evrard, A. E. and Henry, J. P.
Expectations for X-ray cluster observations by the ROSAT
satellite. {\em Astrophys. J.} 383, 95-103, 1991.
\medskip

\item
Finoguenov, A., Briel, U. G. and Henry, J. P.,
XMM-Newton discovery of an X-ray filament in Coma,
{\em Astron. Astrophys.}, 410, 777-784, 2003.
\medskip

\item
Forman, W. and Jones, C.,
X-ray-imaging observations of clusters of galaxies,
{\em Ann. Rev. Astron. Astroph.}, 20, 547-585, 1982.
\medskip

\item
Giovannini, G. and Feretti, L., Halo and relic sources in clusters of galaxies,
{\em New Astronomy}, 5, 335-347, 2000
\medskip

\item
Jones, F.C., and D.C. Ellison,  The plasma physics of shock
acceleration, {\em Space Sci. Rev.}, 58, 259-346, 1991.
\medskip

\item
Kaastra, J. S., Tamura, T., Peterson, J. R., Bleeker, J. A. M. et al.
 Spatially resolved X-ray spectroscopy of cooling clusters of galaxies,
{\em Astron. Astrophys.}, 413, 415-439, 2004.
\medskip

\item
Kaiser, N., Evolution and clustering of rich clusters, {\em MNRAS}
222, 323-345, 1986.
\medskip

\item
Kan, J. R., Lyu, L.H., Mandt, M.E.,  Quasi-parallel collisionless
shocks , {\em Space Sci. Rev.}, 57, 201-236, 1991.
\medskip

\item Kang, H.\, Jones, T.W., Gieseler, U.D.J.,
Numerical studies of cosmic-ray injection and acceleration
, {\em Astrophys. J.}, 579 , 337-358, 2002.
\medskip

\item
Kennel, C. F., Edmiston,J.P., Hada, T.,   A quarter century of
collisionless shock research, in: Collisionless shocks in
heliosphere: A tutorial review, eds. R. G. Stone and
B.T.Tsurutani, pp. 1-36, AGU, Washington, DC, 1985.
\medskip

\item
Knight, P. A.,  Ponman, T. J., The properties of the hot gas
in galaxy groups and clusters from 1D hydrodynamical simulations -
I. Cosmological infall models, {\em MNRAS} 289, 955-972, 1997.
\medskip

\item Malkov, M.A. and Drury, L.O'C.,   Nonlinear theory of diffusive
acceleration  of particles by shock waves, {\em Rep. Progr.
Phys.}, 64, 429-481, 2001.
\medskip

\item
Markevitch, M., The Lx-T relation and temperature function for
nearby clusters revisited , {\em Astrophys. J.}, 504, 27-34, 1998.
\medskip

\item
Markevitch, M.,Ponman, T. J., Nulsen, P. E. J. et al. Chandra
observation of Abell 2142: Survival of dense subcluster cores in a
merger, {\em Astrophys. J.}, 541, 542-549, 2000.
\medskip

\item McKenzie, J.F., and  V\"olk, H.J., Non-linear theory of cosmic
ray shocks including self-generated \alf-Waves, {\em Astron.
Astrophys.}, 116, 191-200, 1982.
\medskip

\item Mewe, R., Ionization of hot plasmas,
in: Physical processes in hot cosmic plasmas, W.Brinkmann et.al.
(eds.),
 pp. 39 - 64, Kluwer, 1990.
\medskip

\item
Miniati, F., Ryu, D., Kang, H., Jones, T.W., Ostriker, J.P.,
Properties of cosmic shock waves in large-scale structure
formation, {\em Astrophys. J.}, 542, 608-621, 2000.
\medskip

\item
Mohr, J.J., Mathiesen, B., \& Evrard, A.E., Properties of the
intracluster medium in an ensemble of nearby galaxy clusters, {\em
Astrophys. J.}, 517, 627-649, 1999.
\medskip

\item Newman, W.I.\, A.I. Newman, Y.Rephaeli,
Quantification of uncertainty in the measurement of magnetic
fields in clusters of galaxies , {\em Astrophys. J.}, 575,
755-763, 2002.
\medskip

\item Norman, C.A.,  Melrose, D.B., , Achterberg, A.,
The origin of cosmic rays above 10$^{18.5}$ eV , {\em Astrophys.
J.}, 454 , 60-68, 1995.
\medskip

\item Ostriker, J.P., and  Souradeep, T.,
The current status of observational cosmology , astro-ph/0409131,
1-13, 2004.
\medskip

\item
Piffaretti, R., Jetzer, Ph., Kaastra, J. S., Tamura, T.,
Temperature and entropy profiles of nearby cooling flow clusters
observed with \xmm, astro-ph/0412233, 2004, ({\em Astron.
Astrophys.} in press).
\medskip

\item
Quest, K.B., Theory and simulations of collisionless parallel
shocks, {\em J. Geophys. Res.}, 93, 9649-9666, 1988.
\medskip

\item
Rakowski, C.E., Ghavamian, P., and Hughes, J.P., The physics of supernova remnant blast waves. II. electron-ion equilibration in DEM L71 in the
Large Magellanic Cloud, {\em Astrophys. J.}, 590, 846-857, 2003.
\medskip

\item
Raymond, J.C., Optical and UV diagnostics of supernova remnant shocks,
{\em Space Sci. Rev.}, 99, 209- 218, 2001.
\medskip

\item
Reimer, O., Pohl, M., Sreekumar, P., Mattox, J. R.,
 EGRET upper limits on the high-energy gamma-ray emission of galaxy clusters,
{\em Astrophys. J.}, 588, 155-164, 2003.
\medskip

\item
 Rosati, P., Borgani, S., Norman, C.,
The evolution of X-ray clusters of galaxies, {\em Ann. Rev.
Astron. Astroph.}, 40, 539-577, 2002.
\medskip

\item
Ryu, D., Kang, H., Hallman, E., Jones, T.W., Cosmological shock
waves and their role in the large-scale structure of the Universe,
{\em Astrophys. J.}, 593, 599-510, 2003.
\medskip

\item Sanderson, A.J.R., Ponman, T.J., Finoguenov, A., Lloyd-Davies, E.J., Markevitch, M.
 The Birmingham-CfA cluster scaling project - I. Gas fraction and $M-T_x$
 relation, {\em MNRAS}, 340, 989-1010, 2003.
\medskip

\item
Sarazin, C.L., X-ray emission from clusters of galaxies,  Cambridge University
Press, Cambridge, 1988.
\medskip

\item
Schwartz, S.J., M.F. Thomsen, S.J. Bame, et al.,
 Electron heating and the potential jump across fast mode shocks,
{\em J Geophys. Res.}, 93, 12923-12931, 1988.
\medskip

\item Tozzi, P., Norman, C.,The evolution of X-Ray clusters and the entropy of the intracluster medium,
{\em Astrophys. J.}  546, 63 -84, 2001.
\medskip

\item
Tsurutani, B.T. and  Lin, R.P., Acceleration of $>$ 47 keV and $>$
2 keV electrons by interplanetary shocks at 1 AU , {\em J Geophys.
Res.}, 90, 1-11, 1985.
\medskip

\item
 Vikhlinin, A., Markevitch, M. and Murray, S.S.
 A moving cold front in the intergalactic medium of A3667,
 {\em Astrophys. J.} 551, 160-171, 2001.
\medskip

\item  Voit, G.M., Balogh, M.L., Bower, R.G.,
Lacey, C.G., Bryan, G.L. On the origin of intracluster entropy,
{\em Astrophys. J.} 593, 272-290, 2003.
\medskip

\item
White, D.A., Jones, C., Forman, W. An investigation of cooling
flows and general cluster properties from an X-ray image
deprojection analysis of 207 clusters of galaxies, {\em MNRAS},
292 , 419-467, 1997.
\medskip

\item Wu, X-P., Xue, Y-J., M. and Fang, L.Z.
 The $L_{\rm X}-T$ and $L_{\rm X}-\sigma$ relationships for galaxy clusters
 revisited, {\em Astrophys. J.} 524, 22-30, 1999.
\medskip

\end{list}
\end{document}